\documentclass[journal,twocolumn]{IEEEtran}   
\usepackage{graphicx}
\usepackage{epstopdf}
\usepackage{amssymb}
\usepackage{amsfonts}
\usepackage{amsmath}
\usepackage{algorithm}
\usepackage{algorithmic}
\usepackage{subeqnarray}
\usepackage{cases}
\usepackage{bm}
\usepackage{stfloats}

\usepackage{float}
\usepackage{graphicx}  
\usepackage{subfig}    
\usepackage{float}     
\usepackage{subfig}
\usepackage{cite}

\ifCLASSINFOpdf
\else
\fi
\begin{document}

\begin{sloppypar}

\title{Phase Optimization for Massive IRS-aided Two-way Relay Network}

\author{Peng Zhang, Xuehui Wang, Siling Feng, Zhongwen Sun, Feng Shu, and Jiangzhou Wang
\thanks{This work was supported in part by the National Natural Science Foundation of China (Nos. 62071234, 62071289, and 61972093), the Hainan Major Projects (ZDKJ2021022), the Scientific Research Fund Project of Hainan University under Grant KYQD(ZR)-21008 and KYQD(ZR)-21007, and the National Key R\&D Program of China under Grant 2018YFB180110.\emph{(Corresponding authors: Feng Shu)}}
\thanks{Peng Zhang, Xuehui Wang, Siling Feng, Zhongwen Sun, and Feng Shu are  with the School of Information and Communication Engineering, Hainan University,~Haikou,~570228, China.}
\thanks{Jiangzhou Wang is with the School of Engineering and Digital Arts, University of Kent, Canterbury CT2 7NT, U.K. Email: (e-mail: j.z.wang@kent.ac.uk).}
}

\maketitle
\begin{abstract}
In this paper, with the help of an intelligent reflecting surface (IRS), the source (S) and destination (D) exchange information through the two-way decode-and-forward relay (TW-DFR). We focus on the phase optimization of IRS to improve the system sum rate performance. Firstly, a maximizing receive power sum (Max-RPS) method is proposed via eigenvalue decomposition (EVD) with an appreciable sum rate enhancement, which is called Max-RPS-EVD. To further achieve a higher sum rate, a method of maximizing minimum rate (Max-Min-R) is proposed with high complexity. To reduce its complexity, a low-complexity method of maximizing the sum rate (Max-SR) via general power iterative (GPI) is proposed, which is called Max-SR-GPI. Simulation results show that the proposed three methods outperform the random phase method, especially the proposed Max-SR-GPI method is the best one achieving at least 20\% sum rate gain over random phase. Additionally, it is also proved the optimal sum rate can be achieved when TW-DFR and IRS are located in the middle of S and D.
\end{abstract}
\begin{IEEEkeywords}
Intelligent reflecting surface, phase optimization, two-way decode-and-forward relay, general power iterative, rate
\end{IEEEkeywords}
\section{Introduction}
In traditional communication systems, wireless environment is uncontrollable, which will have a great impact on the quality of communication service \cite{zhou_irs,2017Robust,2011Relay,2014Low,2017dm,2021sp}. Due to its ability to intelligently reconfigure the wireless channel environment, intelligent reflecting surface (IRS) has attracted \cite{9136592,wys1,wys2,2022Design}. To achieve an intelligent wireless environment created by IRS, how to combine IRS and existing wireless technology is crucial. IRS has been widely used in different wireless networks to improve the data rate, coverage, security, or energy harvesting in \cite{Wuqq-2021}, such as directional modulation \cite{9530396}, spatial modulation \cite{2022sp,9171362}, secure wireless information and power transfer\cite{9288742},  MIMO\cite{mimo1,mimo2,2022Joint_Precoder_mimo}, unmanned aerial vehicle
(UAV)  communication\cite{IRS_UAV} and relay network \cite{2022irs_relay_Deployment,2021irs_relay_Resource_Allocation,2021irss_df,Beat-Relaying2020,Hybrid-Relay2020,2022wang,Joint-Beamforming2021,2022sun}.

First of all, some work related to two-way relay was summarized. In \cite{2010_Asymptotically}, a cooperative scheme for distributed two-way amplify-and-forward (AF) relays was designed, which is asymptotically optimal in the high signal-to-noise ratio (SNR) region. Through a tight approximation of the SNR expression, an upper bound of the mean square error (MSE) is obtained. Meanwhile, the minimum MSE of transmission in both directions was achieved.
Aiming at minimizing the total power of both transceivers, \cite{2012two_way_Semi} presented a total power minimization method with a semi-closed-form solution, which is subjected to SNR constraints at the two transceivers. The proposed method was applied to optimally find relay beamforming weights and transceiver transmit power. To improve spectral efficiency of one-way relay network, single relay-selection strategies with amplify forwarding and network coding in multi-two-way network were proposed in \cite{2012Single_Relay_Selections}, which could achieve full diversity and low bit error rate.
Under the total power constraint, power allocation among the cognitive nodes has an important influence on performance. An optimal power allocation algorithm and a simple distributed power allocation scheme were considered to allocate radio resources in \cite{2012Optimal_power}, where the minimum signal-to-interference-to-noise ratio at secondary users ($\text{SU}_\text{S}$) was viewed as the objective function, and the total transmit power of cognitive nodes and the primary user (PU) interference threshold were regarded as constraints.
 In order to obtain better bit error rate performance of a two-way decode and forward (DF) relay network, a power allocation method for each node was proposed in \cite{2013Joint_relay}, where a closed expression for the end-to-end bit error rate was derived.

IRS is also viewed as a reflecting relay. Recently, the IRS-aided  networks have emerged \cite{Sum-Rate2020,Two-Way2020,OFDM2020,twohop_irs,2022Performance_Analysis}.
 \cite{Sum-Rate2020} the author  presented an IRS-aided full-duplex two-way system with multi-antenna users, the maximum system sum rate was obtained by jointly optimizing the source precoders and the IRS phase.  In \cite{Two-Way2020},  the outage probability and spectral efficiency were derived for reconfigurable intelligent surface (RIS) by approximating instantaneous signal-to-interference-plus-noise ratio (SINR) as a sum of product of two Rayleigh random variables (RVs).
 Finally, the outage probability decreased at the rate of $(\log(\rho)/\rho)^L$ and the spectral efficiency increased at rate of $\log(\rho)$, where $\rho$ was average SINR.
 In an RIS-aided two-way orthogonal frequency-division multiplexing (OFDM) system with $K$ user pairs \cite{OFDM2020}, the minimum bidirectional weighted sum rate was maximized via semi-definite relaxation and projected sub-gradient methods.
\cite{twohop_irs} formulated a non-convex problem describing the rate distribution and found its solution using successive convex approximation (SCA). On this basis, a distributed low-complexity IRS control algorithm based on alternating direction multiplier method (ADMM) and alternating optimization method (AO) was proposed, aiming at maximizing the reachable sum rate, and the performance of the algorithm was improved. A mathematical framework for RIS assisted networks was presented in \cite{2022Performance_Analysis}, which was applied to statistically describe cascaded channels. Through the analysis of high SNR, the expressions of outage probability and bit error rate were reduced to easy-to-calculate forms, which were used to evaluate the performance. Moreover, internal information for system design was provided by high SNR analysis, which can be extended to multi-RIS networks.

Since IRS is a low cost and low power consumption passive reflector,  combining IRS and relay network can strike a good balance among cost, performance, and coverage.
Recently, there has been some research work on the combination of relay and IRS \cite{2022irs_relay_Deployment,2021irs_relay_Resource_Allocation,2021irss_df,Beat-Relaying2020,Hybrid-Relay2020,2022wang,Joint-Beamforming2021,2022sun}.
The IRS deployment has a substantial impact on capacity performance, so two IRS deployment strategies called single-IRS and multi-IRS were presented in an IRS-assisted decode-and-forward (DF) relay network \cite{2022irs_relay_Deployment}. For multi-IRS deployment, a cooperative IRS passive beamforming method was proposed. Compared with the single-IRS scenario, the method with enough IRS elements can achieve a larger capacity of scaling order.
In order to improve the rate of IRS-aided communication system, a new relaying IRS-assisted network was designed in \cite{2021irs_relay_Resource_Allocation}, where the IRS controller was regarded as a DF relay. To maximize the achievable rate, the time allocations for DF relay and the IRS passive beamforming were jointly optimized. It proved that the rate performance of the proposed new system is better than that of the conventional IRS without relaying.
The authors proposed a multi-IRS-aided DF relay network \cite{2021irss_df}, where the IRS configuration including the numbers of IRSs and IRS reflecting elements was considered to maximize the rate. For a given channel coherence time, it was indicated that fewer IRSs with a larger number of reflecting elements tend to have better rate performance.


In \cite{Beat-Relaying2020}, the authors made a comparison between IRS and one-way decode-and-forward (DF) relay and found that in  that an IRS with hundreds of elements,
might beat DF relaying in terms of energy efficiency.
 \cite{Hybrid-Relay2020} proposed a novel hybrid single-antenna relay and IRS assistance system for future wireless networks,  derived tight upper bounds for the achievable rates, and found in the low and middle SNR regions, using a single-antenna DF relay could save a large-scale number of reflecting elements to achieve the same performance.
In \cite{2022wang}, the authors added the antennas at relay, and proposed three high-performance methods for maximizing received power, which were an alternately iterative structure, null-space projection plus maximum ratio combining (MRC) and IRS element selection plus MRC, respectively. By optimizing beamforming at relay and phase shift (PS) at IRS, the rate performance was obviously improved.
In \cite{Joint-Beamforming2021}, an RIS-aided two-way amplify and-forward (AF) relay network was considered. In RIS-aided multiple-antennas BS, the PS matrix was  obtained by the  SNR-upper-bound-maximization method and
genetic-SNR-maximization  method. In \cite{2022sun}, pilot pattern and channel estimation was investigated in an IRS-assisted two-way relay network to show the fact that the Hadamard matrix is ideal for channel estimation using low-resolution phase shifters.

To the best of our knowledge, there is no literature making an investigation of beamforming in an IRS-aided multi-antenna DF relay network. In this paper,   we focus mainly
on  optimizing the PS of the IRS in such a system. Our main contributions are summarized as follows:

\begin{enumerate}
\item
To enhance the exchange rate between source and destination, an  IRS-aided two-way decode-and-forward relay (TW-DFR) network model is proposed. We focus on optimizing the  IRS PS in the first time slot due to the fact that the IRS phase adjustment can be modelled as a typical three-point IRS-aided network in the second time-slot.
First, a maximizing receive power sum (Max-RPS) method is proposed and solved via eigenvalue decomposition (EVD) with an appreciable sum rate enhancement, which is called Max-RPS-EVD. Compared to the random phase, the proposed Max-RPS-EVD achieves an appreciable sum rate enhancement.
\item
To further improve the  sum rate,  a method of maximizing the minimum rate (Max-Min-R) is developed to address the phase optimization problem. However,  its computational complexity is very high.  Using the rule of maximizing the sum rate (Max-SR), a low-complexity method using general power iterative (GPI) is proposed, called Max-SR-GPI. Simulation results show that the proposed Max-SR-GPI performs better than Max-RPS-EVD, Max-Min-R, and random phase in terms of sum rate, and  at least 20\% sum rate gain over the  random phase as the number of the IRS elements  increases up to 80.
\end{enumerate}

The remainder of this paper is organized as follows.
Section II describes the system model and the sum rate. In Section III, three methods for optimizing the phase are proposed.  Simulation
results and related analysis are presented in Section IV. Finally,
we  our conclusions in Section V.

\emph{Notations}: throughout the paper, scalars, vectors and matrices are respectively represented by  lower case, bold lower case, and bold upper case. $(\cdot)^*$, $(\cdot)^H$, $(\cdot)^\dagger$ stand for matrix conjugate, conjugate transpose, and Moore-Penrose pseudo inverse, respectively.
$\mathbb{E}\{\cdot\}$, $\|\cdot\|$ , $\text{tr}(\cdot)$ and $\textbf{I}_{N+1}$       denote expectation operation, 2-norm , the trace of a matrix and the $(N+1)\times (N+1)$ identity matrix, respectively.
\section{System model}
Fig. 1 sketches an  IRS-aided multi-antenna DF relay network including source (S), relay station (RS), IRS, and destination (D).  S,  IRS, RS, and D are employed 1, $N$, $M$ and 1 antennas. Due to a significant path loss, it is assumed that the power of the signals reflected by the IRS two or more times is negligible. Define $ \textbf{h}_{SI}\in\mathbb{C}^{N\times 1}$,  $\textbf{H}_{IR}\in\mathbb{C}^{M\times N}$, $\textbf{h}_{SR}\in\mathbb{C}^{M\times 1}$, $\textbf{h}_{DI}\in\mathbb{C}^{N\times 1}$,  $\textbf{h}_{DR}\in\mathbb{C}^{M\times 1}$, $\textbf{H}^H_{RI}\in\mathbb{C}^{N\times M}$, $\textbf{h}_{IS}^H\in\mathbb{C}^{1\times N}$,  $\textbf{h}_{ID}^H\in\mathbb{C}^{1\times N}$, $\textbf{h}_{RS}^H\in\mathbb{C}^{1\times M}$, and $\textbf{h}_{RD}^H\in\mathbb{C}^{1\times M}$ as the channels from  S to IRS,  I to  RS, S to RS, D to IRS, D to RS, RS to IRS, IRS to  S, IRS to D, RS to  S,
RS to D, respectively.

In the first time slot, the received signal at RS is
\begin{align}\label{yr_111}
\textbf{y}_{RS}=&\sqrt{P_{S}}\textbf{h}_{SR}x_{S}+\sqrt{P_{S}}\textbf{H}_{IR}\bm\Theta\textbf{h}_{SI}x_{S}+\nonumber\\
&~~~~~\sqrt{P_{D}}\textbf{h}_{DR}x_{D}+\sqrt{P_{D}}\textbf{H}_{IR}\bm\Theta\textbf{h}_{DI}x_{D}+\textbf{w}_{R}\nonumber\\
                 =&\sqrt{P_{S}}\textbf{h}_{SIR}x_{S}+\sqrt{P_{D}}\textbf{h}_{DIR}x_{D}+\textbf{w}_{R}.\nonumber\\
                 =&\sqrt{P_{S}}\textbf{h}_{SR}x_{S}+\sqrt{P_{D}}\textbf{h}_{DR}x_{D}+\nonumber\\
&~~~~~\underbrace{\textbf{H}_{IR}\bm\Theta(\sqrt{P_S}\textbf{h}_{SI}x_{S}+\sqrt{P_D}\textbf{h}_{DI}x_{D})}_{r_\Theta}+\textbf{w}_{R}.
\end{align}
where
\begin{subequations}
\begin{align}
\textbf{h}_{SIR}=&\textbf{h}_{SR}+\textbf{H}_{IR}\bm\Theta\textbf{h}_{SI},
\end{align}
\begin{align}
\textbf{h}_{DIR}=&\textbf{h}_{DR}+\textbf{H}_{IR}\bm\Theta\textbf{h}_{DI},
\end{align}
\end{subequations}
 $P_{S}$ and $P_{D}$ are the transmit power at S and D, $x_{S}$ is the date symbol at S with $\mathbb{E}[|x_{S}|^{2}]=1$, and $x_{D}$ is the date symbol at D with $\mathbb{E}[|x_{D}|^{2}]=1$. It is assumed that $x_{S}$ and  $x_{D}$ are independent signals, and obey the same distribution $\mathcal{CN}(0, 1)$, therefore, $\mathbb{E}[|x_{S}x_{D}^H|^{2}]=0$.  $\textbf{w}_{R}\in\mathbb{C}^{M\times 1}$, and $\textbf{w}_{R}\sim\mathcal{CN}(\textbf{0}, \sigma_{R}^{2}\textbf{I}_{M})$
denote the complex additive white Gaussian noise (AWGN)
vectors at RS receiver.   In (\ref{yr_111}), $\bm\Theta$ is the PS matrix of IRS defined as  $\bm\Theta=\text{diag}(e^{j\theta_{1}},...,e^{j\theta_{N}})$, where $\theta_{i}\in[0,2\pi)$ is the PS for the $i$th reflecting element, respectively.
\begin{figure}[h]
\centering
\includegraphics[width=0.5\textwidth,height=0.28\textheight]{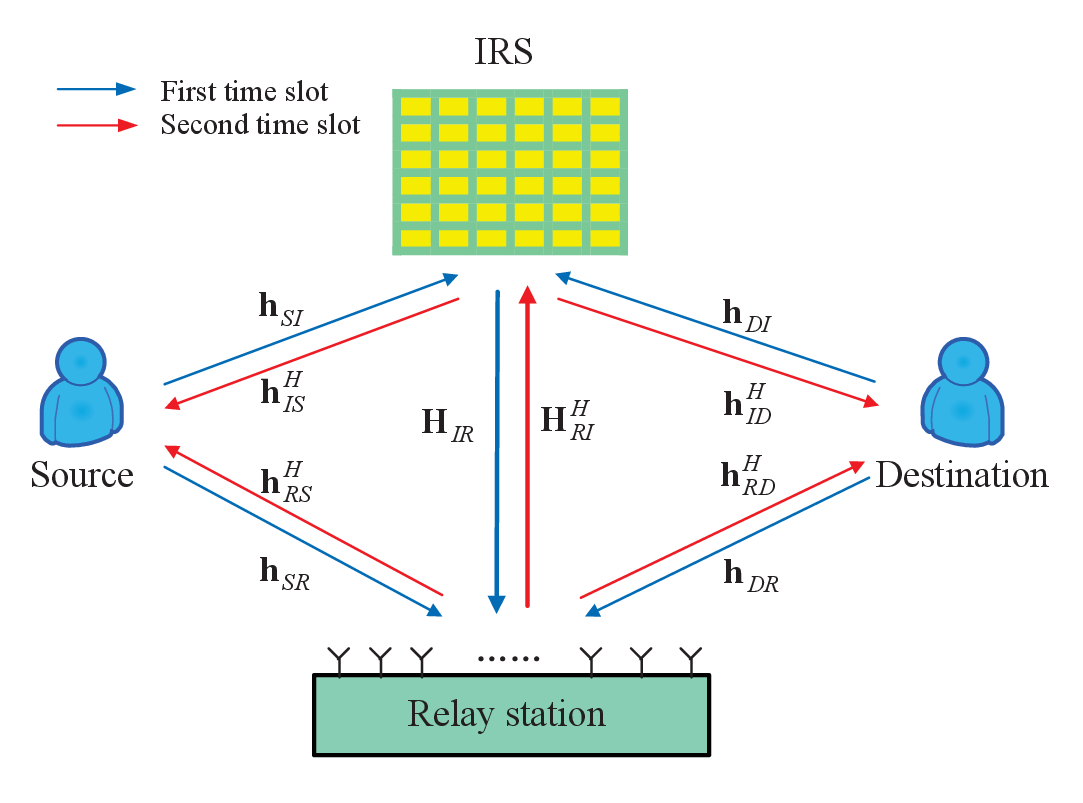}
\centering
\setlength{\abovecaptionskip}{0pt}
\setlength{\belowcaptionskip}{0pt}
\caption{System model for an IRS-aided TW-DFR network.}
\label{fig:label}
\end{figure}

It is assumed that RS decodes  successfully $x_{S}$ and $x_{D}$ by using some traditional decoding methods like sphere decoding, maximum likelihood etc.  Then, RS transmits the network coding symbols $\tilde{x}_S$ and $\tilde{x}_D$ to S and D, respectively. Then, the received signals
\begin{equation}
\begin{aligned}
y_{S}=
\sqrt{P_{R}}\textbf{h}_{RS}^H\tilde{\textbf{x}}_D+\sqrt{P_{R}}\textbf{h}_{IS}^H\bm\Psi \textbf{H}_{RI}^H\tilde{\textbf{x}}_D+w_{S},
\end{aligned}
\end{equation}
\begin{equation}
\begin{aligned}
y_{D}=
\sqrt{P_{R}}\textbf{h}_{RD}^H\tilde{\textbf{x}}_S+\sqrt{P_{R}}\textbf{h}_{ID}^H\bm\Psi \textbf{H}_{RI}^H\tilde{\textbf{x}}_S+w_{D}.
\end{aligned}
\end{equation}

It is assumed that the channels are reciprocity between the first and second time slots, we have $\textbf{h}_{RD}^H=\textbf{h}_{DR}^H, ~\textbf{h}_{ID}^H=\textbf{h}_{DI}^H, ~\textbf{H}_{RI}^H=\textbf{h}_{IR}^H,~ \textbf{h}_{RS}^H=\textbf{h}_{SR}^H,~  \textbf{h}_{SI}^H=\textbf{h}_{SI}^H$. Therefore, the above formulas can be rewritten as
\begin{equation}
\begin{aligned}\label{ys}
y_{S}=
\sqrt{P_{R}}\textbf{h}_{SR}^H\tilde{\textbf{x}}_D+\sqrt{P_{R}}\textbf{h}_{SI}^H\bm\Psi \textbf{H}_{IR}^H\tilde{\textbf{x}}_D+w_{S},
\end{aligned}
\end{equation}
\begin{equation}
\begin{aligned}\label{yd}
y_{D}=
\sqrt{P_{R}}\textbf{h}_{DR}^H\tilde{\textbf{x}}_S+\sqrt{P_{R}}\textbf{h}_{DI}^H\bm\Psi \textbf{H}_{IR}^H\tilde{\textbf{x}}_S+w_{D}.
\end{aligned}
\end{equation}
where $P_{R}$ is the transmit power at RS, $w_{S}$  is the AWGN  with zero mean and variance of $\sigma_{S}^{2}$ at S, $w_{D}$  is the AWGN  with zero mean and variance of $\sigma_{D}^{2}$ at D, and
$\bm\Psi=\text{diag}(e^{j\psi_{1}},...,e^{j\psi_{N}})$ is the PS of matrix IRS  during the second time slot. The system sum rate  is given by
\begin{align}\label{rate}
R_{IR}(\bm\Theta)=
&\frac{1}{2}\{\min \{R_{SIR},R_{RID}\}+\min\{R_{DIR},R_{RIS}\}\}.
\end{align}
where $R_{SIR}$ and $R_{DIR}$ respectively denote the rates from S to RS and D to RS in the first time slot. $R_{RIS}$ and $R_{RID}$ respectively represent the rates from RS to S and RS to D in the second time slot.
\begin{equation}\label{rate1}
\begin{aligned}
R_{SIR}=\text {log}_{2}\left(1+\frac{P_S\textbf{h}_{SIR}^H\textbf{h}_{SIR}}{\sigma_R^2}\right),
\end{aligned}
\end{equation}
\begin{equation}\label{rate2}
\begin{aligned}
R_{DIR}=\text
{log}_{2}\left(1+\frac{P_D\textbf{h}_{DIR}^H\textbf{h}_{DIR}}{\sigma_R^2}\right),
\end{aligned}
\end{equation}
\begin{equation}
\begin{aligned}
R_{RID}=\text
{log}_{2}\left(1+\frac{P_R\textbf{h}_{RID}^H\textbf{h}_{RID}}{\sigma_D^2}\right),
\end{aligned}
\end{equation}
\begin{equation}
\begin{aligned}
R_{RIS}=\text
{log}_{2}\left(1+\frac{P_R\textbf{h}_{RIS}^H\textbf{h}_{RIS}}{\sigma_S^2}\right).
\end{aligned}
\end{equation}
\section{Proposed  methods of optimizing the IRS PS }
In  (\ref{rate}),  PS has a significant impact on the performance of proposed IRS-aided TW-DFR network. In this section, we  focus on PS optimization. In the first time slot, three high-performance methods, namely Max-RPS-EVD, Max-Min-R and Max-SR-GPI, are proposed to optimize PS in order to achieve an improved sum rate performance. In the second time slot, the system model in Fig.~1 can reduce to a typical three-point IRS-aided network, which may be addressed similarly to \cite{Joint-Beamforming2021}.
\subsection{Proposed  Max-RPS-EVD}
From (\ref{yr_111}), we have
\begin{equation}
\begin{aligned}
\mathbb{E}(r_\Theta^Hr_\Theta)
=&\bm\theta^H\textbf{H}\bm\theta.
\end{aligned}
\end{equation}
where
\begin{subequations}
\begin{equation}
\begin{aligned}
\text {diag}({\textbf{h}}_{SI})\bm\theta=\bm\Theta{\textbf{h}}_{SI},
\end{aligned}
\end{equation}
\begin{align}
\text {diag}({\textbf{h}}_{DI})\bm\theta=\bm\Theta{\textbf{h}}_{DI},
\end{align}
\begin{equation}
\begin{aligned}
\textbf{H}=&P_{S}\text {diag}({\textbf{h}}_{SI})^H\textbf{H}_{IR}^H\textbf{H}_{IR} \text {diag}({\textbf{h}}_{SI})+\\
&P_{D}\text {diag}({\textbf{h}}_{DI})^H\textbf{H}_{IR}^H\textbf{H}_{IR} \text {diag}({\textbf{h}}_{DI}).\\
\end{aligned}
\end{equation}
\end{subequations}
 the first term of the right-hand side of (\ref{yr_111}) is independent of $\bm\Theta$, the optimization problem of  maximizing receive power  is recasted as follows:
\begin{subequations}\label{theta}
\begin{align}
&\max \limits_{\bm{\theta}}
~~~\bm\theta^H\textbf{H}\bm\theta~~~~~~~~\\
&~\text{s.~ t. }~\|\bm{\theta}(n)\|^2=1, \forall n.&
\end{align}
\end{subequations}
which is relaxed to
\begin{subequations}\label{new-relax-theta}
\begin{align}\label{theta}
&\max \limits_{\bm{\theta}}
~~~\bm\theta^H\textbf{H}\bm\theta~~~~~~~~\\
&~\text{s.~ t. }~\|\bm{\theta}\|^2=N.&
\end{align}
\end{subequations}
and gives the associated Lagrangian
\begin{equation}
\begin{aligned}
 L(\bm{\theta},\lambda)&=\bm\theta^H\textbf{H}\bm\theta+\lambda(\|\bm{\theta}\|^{2}-N).
\end{aligned}
\end{equation}
Setting the partial derivative of the Lagrangian function with respect to $\bm{\theta}^*$ with zero yields
\begin{equation}
\begin{aligned}
&\frac{\partial L(\bm{\theta},\lambda)}{\partial \bm{\theta}^*}=\textbf{H}\bm\theta+\lambda\bm\theta=0,
\end{aligned}
\end{equation}
Clearly, the optimal  $\bm\theta$ is the eigenvector of matrix $\textbf{H}$.  To maximize the  objective function in (\ref{new-relax-theta}) means $\bm\theta$ is the the eigenvector $\lambda_1$ corresponding to the maximum eigenvalue $\textbf{u}_1$ as the solution to equation  (\ref{theta}).  To obtain the eigenvector, the eigenvalue decomposition of \textbf{H} is performed as $\textbf{H}=\textbf{U}\bm\Sigma\textbf{U}^H$, where $\textbf{U}=[\textbf{u}_1, \cdots, \textbf{u}_M], \bm\Sigma=\text{diag}(\lambda_1, \cdots, \lambda_M)$.
we can get
\begin{align}
\bm\theta=e^{j\arg~\textbf{u}_1} .
\end{align}
\subsection{Proposed Max-Min-R}
In the preceding subsection, the proposed Max-RPS-EVD can maximize the receive power sum, but it is very difficult for it to make a good  balance between source and destination. It is possible that the method enhances the S-IRS-RS signal, and compresses  the D-IRS-RS signal. In what follows, to achieve a fairness between source and destination,  a new method Max-Min-R is presented.

Defining a new optimization variable $\overline{\bm\theta}=[\bm\theta^H,1]^H$,
 channel matrices $\textbf{h}_{SIR}$ and $\textbf{h}_{DIR}$ can be rewritten as
\begin{flalign}
\textbf{h}_{SIR}=&\textbf{h}_{SR}+\textbf{H}_{IR}\bm\Theta\textbf{h}_{SI}\nonumber\\
=&\textbf{h}_{SR}+\textbf{H}_{IR}\text{diag}(\textbf{h}_{SI})\bm\theta\nonumber\\
=&[\textbf{H}_{IR}\text{diag}(\textbf{h}_{SI}),\textbf{h}_{SR}][\bm\theta^H,1]^H\nonumber\\
\triangleq&\overline{\textbf{H}}_{SIR}\overline{\bm\theta},
\end{flalign}
\begin{flalign}
\textbf{h}_{DIR}=&\textbf{h}_{DR}+\textbf{H}_{IR}\bm\Theta\textbf{h}_{DI}\nonumber\\  =&\textbf{h}_{DR}+\textbf{H}_{IR}\text{diag}(\textbf{h}_{DI})\bm\theta\nonumber\\
=&[\textbf{H}_{IR}\text{diag}(\textbf{h}_{DI}),\textbf{h}_{DR}][\bm\theta^H,1]^H\nonumber\\
\triangleq&\overline{\textbf{H}}_{DIR}\overline{\bm\theta}.
\end{flalign}
where
\begin{subequations}
\begin{flalign}
&~~~~~~~~~~~~~~~\overline{\textbf{H}}_{SIR}=[\textbf{H}_{IR}\text{diag}(\textbf{h}_{SI}),\textbf{h}_{SR}],&
\end{flalign}
\begin{flalign}
&~~~~~~~~~~~~~~~\overline{\textbf{H}}_{DIR}=[\textbf{H}_{IR}\text{diag}(\textbf{h}_{DI}),\textbf{h}_{DR}],&
\end{flalign}
\end{subequations}
Using the above definitions,   $R_{SIR}$ and $R_{DIR}$  in (\ref{rate1}) and (\ref{rate2}) can be represent as

\begin{equation}\label{r_sr}
\begin{aligned}
R_{SIR}(\overline{\bm\theta})=\text{log}_2\left(1+\frac{P_S\text{tr}(\overline{\bm\theta}^H\overline{\textbf{H}}_{SIR}^H\overline{\textbf{H}}_{SIR}\overline{\bm\theta})}{\sigma_R^2}\right),
\end{aligned}
\end{equation}
\begin{equation}\label{r_dr}
\begin{aligned}
R_{DIR}(\overline{\bm\theta})=\text{log}_2\left(1+\frac{P_D\text{tr}(\overline{\bm\theta}^H\overline{\textbf{H}}_{DIR}^H\overline{\textbf{H}}_{DIR}\overline{\bm\theta})}{\sigma_R^2}\right).
\end{aligned}
\end{equation}
The Max-Min-R optimization problem is casted as follows
\begin{subequations}
\begin{align}
&~\max\limits_{\overline{\bm{\theta}}}~\min
~~\{R_{SIR}(\overline{\bm{\theta}}),~R_{DIR}(\overline{\bm{\theta}})\}\\
&~~\text{s.~ t. } ~~~~~~~~\|\overline{\bm{\theta}}(n)\|^2=1,~\forall n.
\end{align}
\end{subequations}
Defining
\begin{subequations}
\begin{equation}
\begin{aligned}
~~~~t=\min\{R_{SIR}(\overline{\bm{\theta}}),~R_{DIR}(\overline{\bm{\theta}})\},
\end{aligned}
\end{equation}
\begin{equation}
\begin{aligned}
R_{SIR}(\overline{\bm{\theta}})\geq t,
\end{aligned}
\end{equation}
\begin{align}
R_{DIR}(\overline{\bm{\theta}})\geq t.
\end{align}
\end{subequations}
the above optimization problem is recasted as
\begin{subequations}\label{sdp0}
\begin{align}
&~\max\limits_{\overline{\bm{\theta}},t}
~~t\\
&~~\text{s.~ t. } ~\|\overline{\bm{\theta}}(n)\|^2=1,\forall n,\\
\label{rate-sd}&~~~~~~~~~R_{SIR}(\overline{\bm{\theta}})\geq t,\\
&~~~~~~~~~R_{DIR}(\overline{\bm{\theta}})\geq t.
\end{align}
\end{subequations}
Defining
\begin{align}
\overline{\bm\Theta}=\overline{\bm\theta}\bullet\overline{\bm\theta}^H,
\end{align}
The above problem is converted into
\begin{subequations}\label{sdp}
\begin{align}
&~\max\limits_{\overline{\bm{\Theta}},t}
~~t\\
&~~\text{s.~ t. } ~\overline{\bm\Theta}[n,n]=1, n\in\{1,~2,~\cdots,~N+1\},\\
&~~~~~~~~~\text{rank}(\overline{\bm\Theta})=1, \\
&~~~~~~~~~\overline{\bm\Theta}\succeq0,\\
&~~~~~~~~~\text{log}_2\left(1+\frac{P_S\text{tr}(\overline{\bm\Theta}\overline{\textbf{H}}_{SIR}^H\overline{\textbf{H}}_{SIR})}{\sigma_R^{2}}\right)\geq t,\\
&~~~~~~~~~\text{log}_2\left(1+\frac{P_D\text{tr}(\overline{\bm\Theta}\overline{\textbf{H}}_{DIR}^H\overline{\textbf{H}}_{DIR})}{\sigma_R^{2}}\right) \geq t.
\end{align}
\end{subequations}
which is further relaxed to
\begin{subequations}\label{sdp}
\begin{align}
&~\max\limits_{\overline{\bm{\Theta}},t}~~t\\
&~~\text{s.~ t. } ~\overline{\bm\Theta}[n,n]=1,\forall n\in\{1,~2,~\cdots,~N+1\},\\
&~~~~~~~~~\overline{\bm\Theta}\succeq0,\\
&~~~~~~~~~\text{log}_2\left(1+\frac{P_S\text{tr}(\overline{\bm\Theta}\overline{\textbf{H}}_{SIR}^H\overline{\textbf{H}}_{SIR})}{\sigma_R^{2}}\right)\geq t,\\
&~~~~~~~~~\text{log}_2\left(1+\frac{P_D\text{tr}(\overline{\bm\Theta}\overline{\textbf{H}}_{DIR}^H\overline{\textbf{H}}_{DIR})}{\sigma_R^{2}}\right) \geq t,
\end{align}
\end{subequations}
by removing the rank-1 constraint. Obviously, the above  problem is a semidefinite programming problem, which  can be solved efficiently via CVX. The optimal solution of $\overline{\bm\Theta}$, however, is not generally a rank-one matrix. Therefore, after obtaining the optimal $\overline{\bm\Theta}$ from (\ref{sdp}), it is necessary to find a rank-one solution by using the Gaussian randomization procedure as summarized in \cite{Joint-Beamforming2021}. Once the optimal solution $\overline{\bm\theta}^\bigstar$ is obtained, we obtain $\bm\theta^\bigstar$ as
\begin{equation}\label{sdp2}
\begin{aligned}
\theta^{\bigstar}=e^{j \text{arg}\left([\frac{\overline{\bm\theta}}{\overline{\theta}_{N+1}}]_{1:N}\right)}
\end{aligned}
\end{equation}
\subsection{Proposed Max-SR-GPI }
 In the previous subsection, the proposed Max-Min-R method  has the computational complexity $\mathcal{O}\{N^{6}\}$ float-pointed operations (FLOPs), which is of very high complexity as the number of IRS elements tends to large-scale. To reduce the complexity, a low-complexity Max-SR-GPI method is proposed in the following with the sum rate defined as
\begin{equation}
\begin{aligned}
R_{R}(\overline{\bm{\theta}})=&R_{SIR}(\overline{\bm{\theta}})+R_{DIR}(\overline{\bm{\theta}})\\
\end{aligned}
\end{equation}
The optimization problem is formulated as follows:
\begin{subequations}\label{max-sd}
\begin{align}
&~~~~~~~~~~~~\max\limits_{\overline{\bm{\theta}}} ~~R_{R}(\overline{\bm{\theta}})\\
&~~~~~~~~~~~~~\text{s.~ t. } ~\|\overline{\bm{\theta}}(n)\|^2=1,\forall n.&
\end{align}
\end{subequations}
which is equivalent to
\begin{subequations}\label{gpl-max}
\begin{align}
&~~\max\limits_{\overline{\bm{\theta}}} ~~\overline{\bm{\theta}}^H\textbf{A}_S\overline{\bm{\theta}}\bullet\overline{\bm{\theta}}^H\textbf{A}_D\overline{\bm{\theta}}\\
&~~~\text{s.~ t.} ~\|\overline{\bm{\theta}}(n)\|^2=1,\forall n.&
\end{align}
\end{subequations}
where
\begin{subequations}
\begin{equation}
\begin{aligned}
\textbf{A}_S=\frac{\mathbb{\textbf{I}}_{N+1}}{N+1}+\frac{P_S\overline{\textbf{H}}_{SIR}^H\overline{\textbf{H}}_{SIR}}{\sigma_R^2},
\end{aligned}
\end{equation}
\begin{equation}
\begin{aligned}
\textbf{A}_D=\frac{\mathbb{\textbf{I}}_{N+1}}{N+1}+\frac{P_D\overline{\textbf{H}}_{DIR}^H\overline{\textbf{H}}_{DIR}}{\sigma_R^2}.
\end{aligned}
\end{equation}
\end{subequations}
The problem (\ref{gpl-max}) is relaxed to
\begin{subequations}\label{new-sum-sd}
\begin{align}
&~~\max\limits_{\overline{\bm{\theta}}} ~~\overline{\bm{\theta}}^H\textbf{A}_S\overline{\bm{\theta}}\bullet\overline{\bm{\theta}}^H\textbf{A}_D\overline{\bm{\theta}}\\
&~~~\text{s.~ t.} ~\|\overline{\bm{\theta}}\|^2=N+1.&
\end{align}
\end{subequations}
which can be solved through GPI in accordance with \cite{4531911}. The whole procedure is summarized in Algorithm 1. Optimization problem (\ref{new-sum-sd}) can be rewritten to\\
\begin{subequations}
\begin{align}
&~\max\limits_{\overline{\bm{\theta}}} ~~\frac{\overline{\bm{\theta}}^H\textbf{A}_S\overline{\bm{\theta}}\bullet\overline{\bm{\theta}}^H\textbf{A}_D\overline{\bm{\theta}}}{\overline{\bm{\theta}}^H\textbf{I}_{N+1}\overline{\bm{\theta}}\bullet\overline{\bm{\theta}}^H\textbf{I}_{N+1}\overline{\bm{\theta}}}(N+1)^2\\
&~~\text{s.~ t. } ~\|\overline{\bm{\theta}}\|^2=N+1.
\end{align}
\end{subequations}
Since a is $(N+1)^2$ constant, the above optimization problem can be reduce to
\begin{subequations}
\begin{align}
&~\max\limits_{\overline{\bm{\theta}}} ~~\frac{\overline{\bm{\theta}}^H\textbf{A}_S\overline{\bm{\theta}}\bullet\overline{\bm{\theta}}^H\textbf{A}_D\overline{\bm{\theta}}}{\overline{\bm{\theta}}^H\textbf{I}_{N+1}\overline{\bm{\theta}}\bullet\overline{\bm{\theta}}^H\textbf{I}_{N+1}\overline{\bm{\theta}}}\\
&~~\text{s.~ t. } ~\|\overline{\bm{\theta}}\|^2=N+1.
\end{align}
\end{subequations}
\rule[-2pt]{9cm}{0.05em}
\textbf{Algorithm 1}~Proposed general power iterative algorithm\\
\rule[8pt]{9cm}{0.05em}
\setlength{\abovecaptionskip}{15cm}
1) Set initial solution $\overline{\bm{\theta}}_0$, k=1,\\
2) Compute the function matrix $\textbf{A}(\overline{\bm{\theta}}_{k-1}), \textbf{B}(\overline{\bm{\theta}}_{k-1})$,\\
3)$\textbf{y}_{k}=\textbf{B}(\overline{\bm{\theta}}_{k-1})^\dagger\textbf{A}(\overline{\bm{\theta}}_{k-1})\overline{\bm{\theta}}_{k-1}$,\\
4) $\overline{\bm{\theta}}_{k}=\frac{\textbf{y}_{k}}{\|\textbf{y}_{k}\|_2}$,\\
5) If $\|\overline{\bm{\theta}}_{k}-\overline{\bm{\theta}}_{k-1}\|<\kappa$ stop,\\
~~~ ~~~~~   Otherwise set $ k = k + 1 $ and go to step 2).\\
\rule[-2pt]{9cm}{0.05em}
where
\begin{align}
\textbf{B}(\overline{\bm{\theta}}_{k-1})^\dagger=(\textbf{B}(\overline{\bm{\theta}}_{k-1})^H\textbf{B}(\overline{\bm{\theta}}_{k-1}))^{-1}\textbf{B}(\overline{\bm{\theta}}_{k-1})^H,
\end{align}
\begin{align}
\textbf{A}(\overline{\bm\theta})=\overline{\bm\theta}^H\textbf{A}_S\overline{\bm\theta} \textbf{A}_D +\overline{\bm\theta}^H\textbf{A}_D\overline{\bm\theta} \textbf{A}_S,
\end{align}
\begin{align}
\textbf{B}(\overline{\bm\theta})=\overline{\bm\theta}^H\textbf{I}_{N+1}\overline{\bm\theta} \textbf{I}_{N+1} +\overline{\bm\theta}^H\textbf{I}_{N+1}\overline{\bm\theta} \textbf{I}_{N+1}.
\end{align}
In Algorithm 1, $\kappa$ is  the  tolerance  factor  for  terminating the algorithm, Applying the general power iterative algorithm in Algorithm 1.

\subsection{Complexity Analysis}
Since the method adopted in the second time slot is the same as that of the first time slot, the complexity in both time slots is the same. Therefore, only the complexity of the first time slot is calculated in the following.
The complexity of  Max-RPS-EVD FLOPs is
\begin{equation}
\begin{aligned}
&C_{Max-RPS-EVD}=\\
&\mathcal{O}\Big\{N^3+4N^2M+2N^2+4MN+N-1\Big\},
\end{aligned}
\end{equation}
Obviously, the highest order of computational complexity is $N^3$ float-point operations (FLOPs). The complexity of  Max-Min-R is
\begin{equation}
\begin{aligned}
C_{Max-Min-R}=
&\mathcal{O}\Big\{\sqrt{2N+4}\left(3a^3+2ab+bc\right)\Big\}.
\end{aligned}
\end{equation}
where
\begin{align}
a=N^2+2N+2,
\end{align}
\begin{align}
b=N^3+3N^2+2,
\end{align}
\begin{align}
c=3N^2+2MN^2+MN+4N-M+1.
\end{align}
The complexity of  Max-SR-GPI FLOPs is
\begin{equation}
\begin{aligned}
&C_{Max-SR-GPI}=I\mathcal{O}\Big\{\frac{19}{3}N^3+25N^2+\frac{119}{3}N+16+d\Big\},
\end{aligned}
\end{equation}
where $I$ is the average number of iterations, and
\begin{align}
d=2MN^2-N^2+6MN+2M-2N-1.
\end{align}
 Considering the number
of antennas tends to be large-scale or ultra-large-scale, compared
with the Max-Min-R, the computational complexity of the
proposed Max-SR-GPI is significantly reduced.
\section{Simulation and analysis}
\begin{figure}[h]
\centering
\includegraphics[width=0.4\textwidth,height=0.2\textheight]{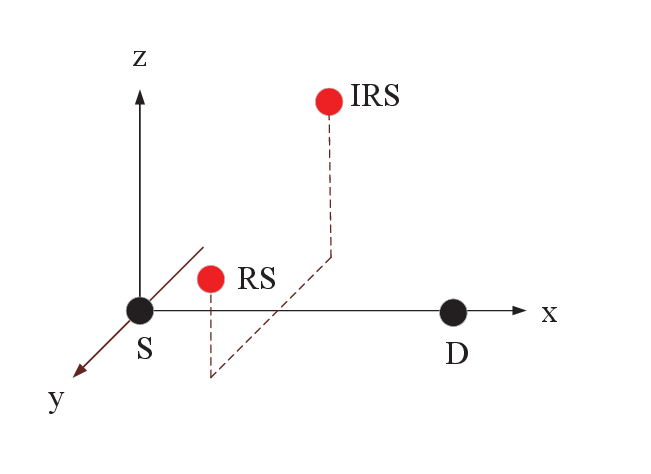}
\centering
\setlength{\abovecaptionskip}{0pt}
\setlength{\belowcaptionskip}{0pt}
\caption{Relative position diagram of S, D, RS and IRS.}
\label{weizhi}
\end{figure}
\begin{figure}[h]
\centering
\includegraphics[width=0.5\textwidth,height=0.3\textheight]{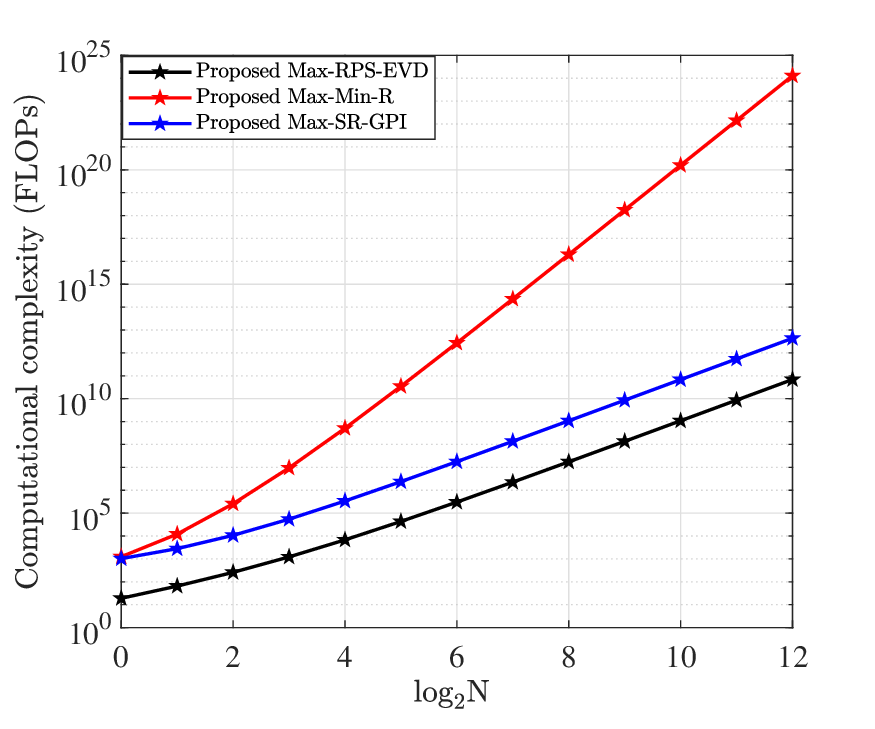}
\centering
\setlength{\abovecaptionskip}{15pt}
\setlength{\belowcaptionskip}{15pt}
\caption{Computational complexity versus  $N$ with $M=$ 2, $\kappa = 10^{-12}$, $I =$ 10, and $d=$ 50m.}
\label{flots}
\end{figure}
\begin{figure}[h]
\centering
\includegraphics[width=0.5\textwidth,height=0.3\textheight]{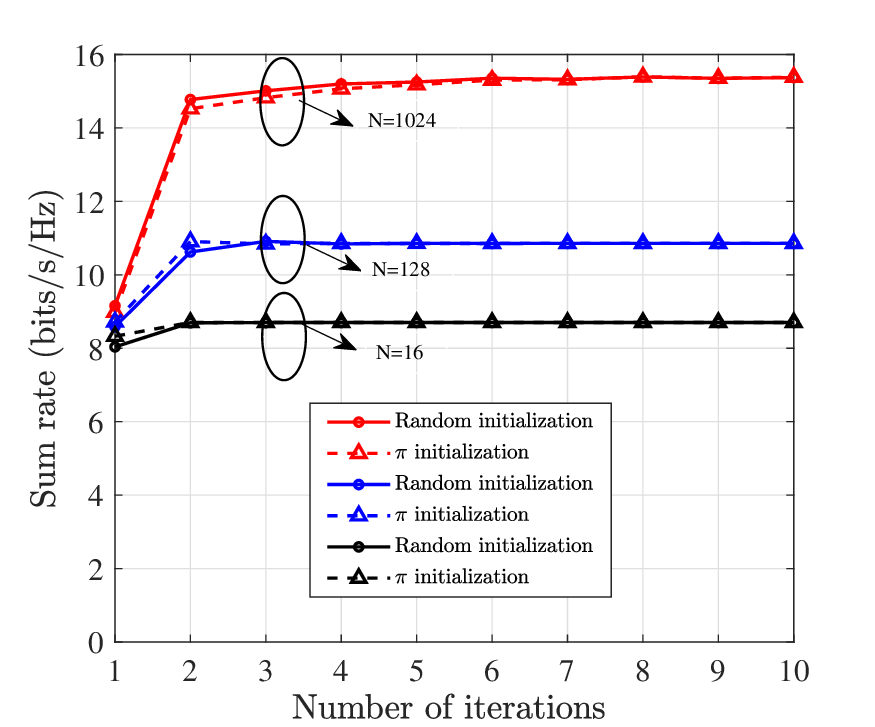}
\centering
\setlength{\abovecaptionskip}{15pt}
\setlength{\belowcaptionskip}{15pt}
\caption{Convergence of Max-SR-GPI with $M=$ 2, $\kappa=10^{-12}$ and $d=$ 50m.}
\label{gpisr}
\end{figure}
\begin{figure}[h]
    \centering
    \setlength{\abovecaptionskip}{15pt}
    \setlength{\belowcaptionskip}{15pt}
    \subfloat[]{          \label{nnnsr}
         \begin{minipage}[c]{1\linewidth}
            \centering
            \includegraphics[width=1\textwidth,height=0.3\textheight]{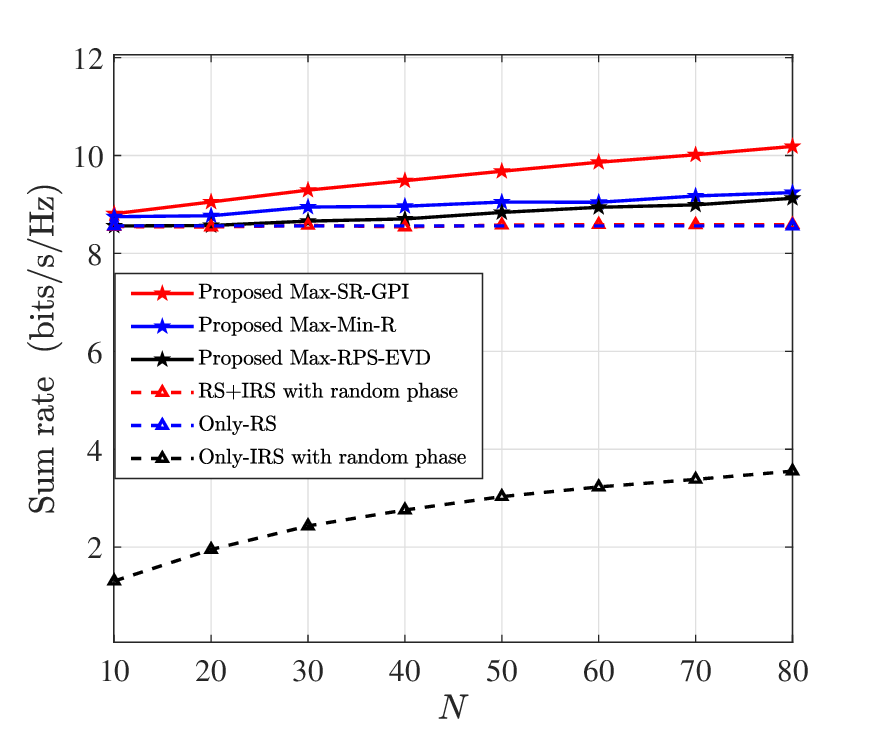}
       \end{minipage}
    }\\
   \subfloat[]{         \label{nnn13456}
         \begin{minipage}[c]{1\linewidth}
         \centering
           \includegraphics[width=1\textwidth,height=0.3\textheight]{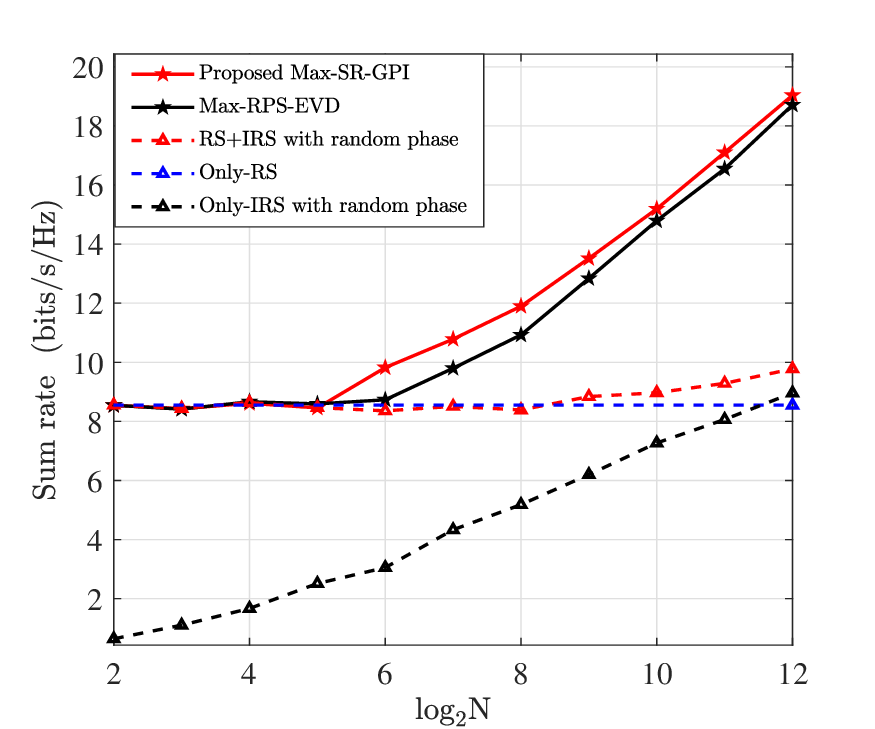}
         \end{minipage}
    }
     \caption{Sum rate versus $N$ with $M=$ 2 and $d=$ 50m, (a) $\kappa=10^{-12}$, (b) $\kappa=10$.}
     \label{nnn}
\end{figure}
\begin{figure}[h]
    \centering
    \setlength{\abovecaptionskip}{15pt}
    \setlength{\belowcaptionskip}{15pt}
    \subfloat[]{          \label{distsr}
         \begin{minipage}[c]{1\linewidth}
            \centering
            \includegraphics[width=1\textwidth,height=0.3\textheight]{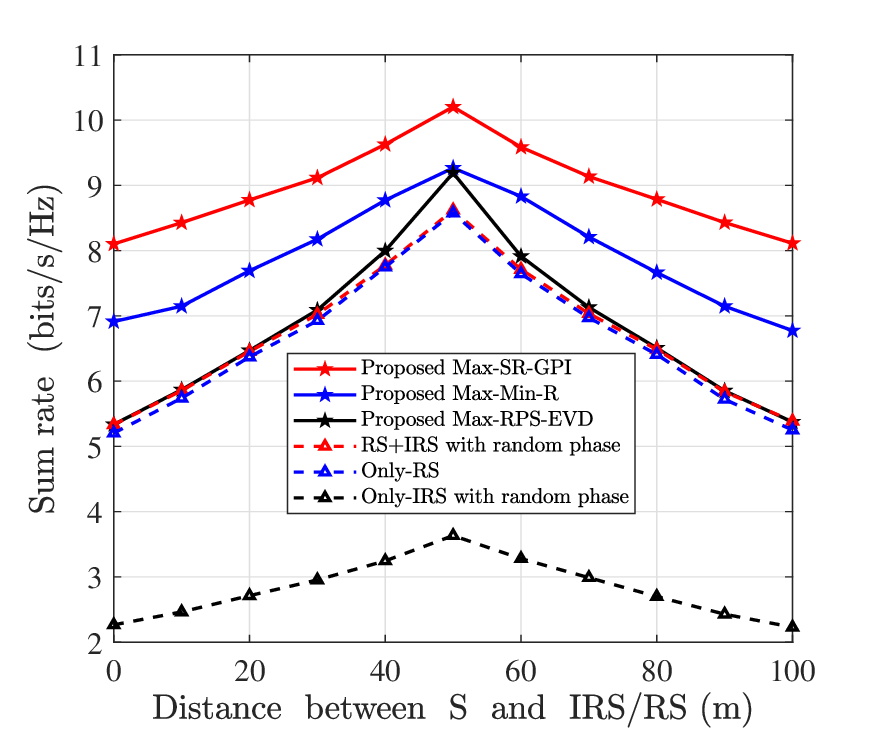}
       \end{minipage}
    }\\
   \subfloat[]{         \label{dist13456}
         \begin{minipage}[c]{1\linewidth}
         \centering
           \includegraphics[width=1\textwidth,height=0.3\textheight]{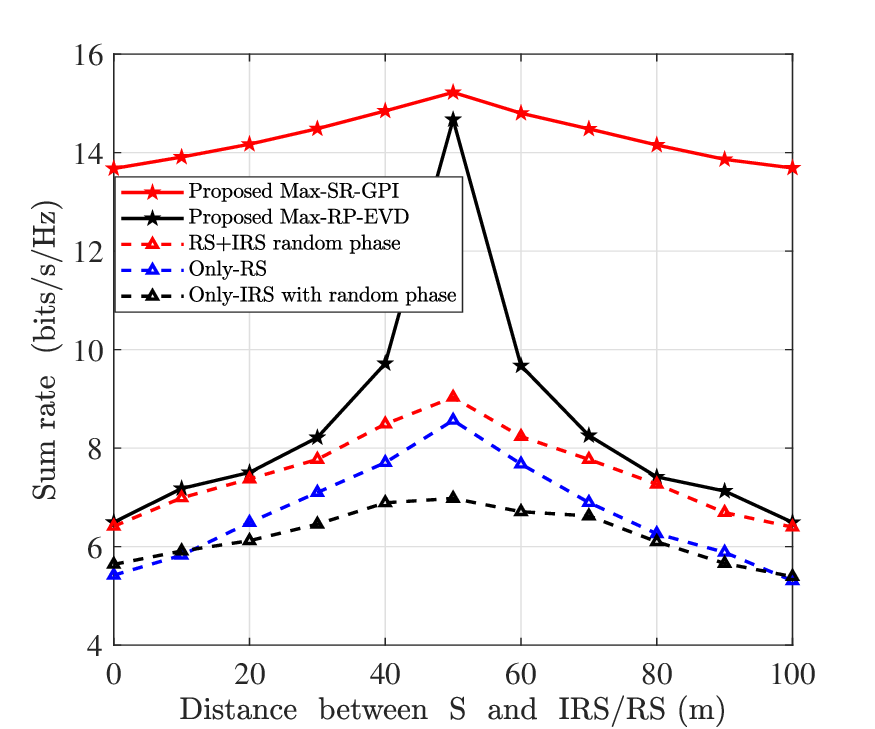}
         \end{minipage}
    }
     \caption{Sum rate versus distance with $M=$ 2, (a) $\kappa = 10^{-12}$ and $N =$ 80, (b) $\kappa =$ 10 and $N =$ 1024.  }
\label{dist}
\end{figure}
\begin{figure}[h]
    \centering
    \setlength{\abovecaptionskip}{15pt}
    \setlength{\belowcaptionskip}{15pt}
    \subfloat[]{          \label{mmmsr}
         \begin{minipage}[c]{1\linewidth}
            \centering
            \includegraphics[width=1\textwidth,height=0.3\textheight]{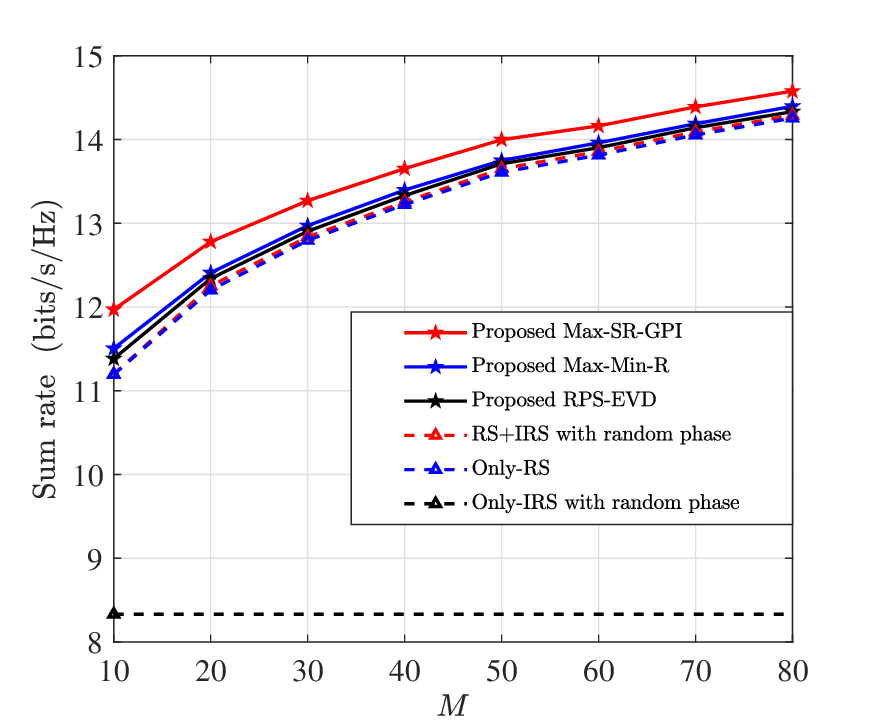}
       \end{minipage}
     }\\
   \subfloat[]{         \label{mmm13456}
         \begin{minipage}[c]{1\linewidth}
         \centering
           \includegraphics[width=1\textwidth,height=0.3\textheight]{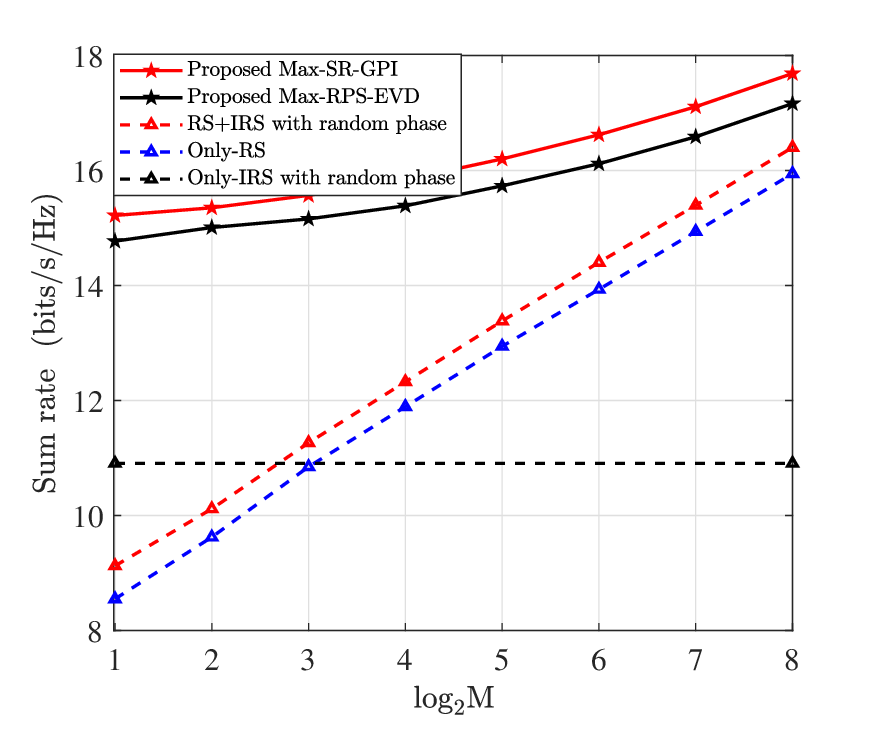}
         \end{minipage}
    }
     \caption{Sum rate versus M with $d=$ 50 ,  (a) $\kappa=10^{-3}$, (b) $\kappa=$ 10.}
\label{mmm}
\end{figure}
In this section, simulation results are provided evaluate  the effectiveness of the proposed methods. As shown in Fig.\ref{weizhi}, we consider a three-dimensional (3D) scenario,  where S, D, IRS and R are located at (0m, 0m, 0m), (100m, 0m, 0m), ($d$, $-$10m, 30m) and ($d$, 10m, 10m), $d \in $[0m, 100m].
The Rayleigh fading model and the path loss at distance d is modeled as $PL(d)=PL_0-10\alpha\text{log}_{10}(\frac{d}{d_0})$, where $ PL_0$ = $-$30dB denotes the path loss at the reference distance $d_0$ = 1m, $\alpha$ denotes the path loss exponent. Specifically, the path loss exponents of S-R, D-R, S-IRS, D-IRS and IRS-R channels
are set to be 3.5, 3.5, 2.5, 2.5 and 2.5, respectively.
Unless specified otherwise, the simulation parameters are set as: $ P_S=P_D=P_R=1$W, $\sigma^2_S=\sigma^2_D=\sigma^2_R=-84\text{dBm}$. For comparison, the benchmark schemes are given as follow: (1) RS+IRS with random phase: The PS of IRS are set randomly in $[0,2\pi)$. (2) Only-RS: There is no use of the
IRS. (3) Only-IRS with random phase: There is no use of the RS, the phase shifts of IRS are random.

Fig. \ref{flots} plots the  curves of computational complexity versus
$N$ with $M$ = 2. It is demonstrated that the computational
complexities of the proposed three methods, Max-RPS-EVD, Max-Min-R and Max-SR-GPI, increase as $N$ increases. Obviously, the second method has the highest computational complexities, which is much higher than that of the third method.

Fig. \ref{gpisr} plots the  proposed Max-SR-GPI for three different sizes of IRS: $N=\{16, 128, 1024\}$ versus the number of iterations.  As the size of IRS increases, it can be seen that the  convergence sum rate becomes slow gradually. For the large-scale case ($N$ = 1024), about six iterations are required to achieve the convergence of the proposed Max-SR-GPI. As the number of elements of IRS varies from 16 to 1024,  the sum rate is doubled approximately. It is verified that initial phase has  a slight impact on the number of iterations, so a random initialization phase is chosen in the following simulation.

Fig. \ref{nnn}  illustrates the transmit  rate versus $N$ for the proposed three methods with random phase as a performance reference at $d = $ 50m.  As the number of the IRS elements increases,  the performance of all methods Max-SR-GPI increase gradually. In particular, the proposed Max-SR-GPI achieves obvious enhancements over Max-Min-R, Max-RPS-EVD and random phase. It demonstrates that the proposed Max-SR-GPI and Max-RPS-EVD are suitable for large-scale IRS, in particular, more satisfactory sum rate can be obtained by Max-SR-GPI.

Fig. \ref{dist}  demonstrates the sum rate versus distance between S and IRS/RS for the proposed methods with random phase as a performance reference.  Clearly, when  IRS and RS are  between source and destination, the proposed methods can achieve their largest rates. Obviously, the performance of Max-SR-GPI  is the best one among three proposed methods, and the rate performance of Max-Min-R is larger than that of Max-RPS-EVD. They perform better than random phase. The proposed method Max-SR-GPI achieves at least 20\% rate gain over  random phase. As the number of IRS reflecting elments increases, the rate gain achieved by the proposed Max-SR-GPI increases.

Fig. \ref{mmm}  demonstrates the sum rate versus $M$ for the proposed methods with random phase as a performance reference.  Fig. \ref{mmm} shows the sum rate versus M for the proposed methods with random phase as a performance benchmark. It can be seen that as the number of RS antennas increases, the sum rate also increases gradually. Clearly, the Max-SR-GPI method is the best one among the proposed three methods.

\section{Conclusions}
In this paper, an IRS-aided two-way decode-and-forward relay (TW-DFR) network model was  presented. Three efficient phase optimization methods: Max-RPS-EVD, Max-Min-R and Max-SR-GPI, have been proposed to enhance the rate performance. Simulation results showed, that the proposed three schemes outperform random phase method in terms of data rate. It is attractive that the proposed Max-SR-GPI method achieves the best rate performance with lower complexity and  at least 20\% rate improvement over random phase, and with the increase of IRS reflecting elements, the improvement is more obvious.

\ifCLASSINFOpdf
\else
\fi

\bibliographystyle{IEEEtran}
\bibliography{newref}

\end{sloppypar}
\end{document}